# Predicting toxicity by quantum machine learning

Teppei Suzuki[1,2] and Michio Katouda[3]

[1] TerraSky Co., Ltd., Taiyo Life Nihombashi Building 15–17F, 2-11-2 Nihombashi, Chuo-ku, Tokyo 103-0027, Japan

[2] Quemix, Inc., Taiyo Life Nihombashi Building 16F, 2-11-2 Nihombashi, Chuo-ku, Tokyo 103-0027, Japan

[3] Research Organization for Information Science and Technology, Sumitomo-Hamamatsucho Building 7F, 1-18-16 Hamamatsucho, Minato-ku, Tokyo 105-0013, Japan

**E-mail:** tsuzuki@quemix.com

**Abstract**

In recent years, parameterized quantum circuits have been regarded as machine learning models within the framework of the hybrid quantum–classical approach. Quantum machine learning (QML) has been applied to binary classification problems and unsupervised learning. However, practical quantum application to nonlinear regression tasks has received considerably less attention. Here, we develop QML models designed for predicting the toxicity of 221 phenols on the basis of quantitative structure activity relationship. The results suggest that our data encoding enhanced by quantum entanglement provided more expressive power than the previous ones, implying that quantum correlation could be beneficial for the feature map representation of classical data. Our QML models performed significantly better than the multiple linear regression method. Furthermore, our simulations indicate that the QML models were comparable to those obtained using radial basis function networks, while improving the generalization performance. The present study implies that QML could be an alternative approach for nonlinear regression tasks such as cheminformatics.

## 1. Introduction

Quantitative structure activity relationship (QSAR) is one of major computational molecular modeling methods. The QSAR approach attempts to correlate molecular descriptors of compounds with their physicochemical properties; over the past decades, it has been used for predicting toxicity and bioactivities as well as finding new drug leads in chemical and pharmaceutical areas [1–4]. Nowadays, owing to rapid development of information and communication technologies, huge amounts of physicochemical data coming from a variety of resources have been accumulated. Currently, databases containing millions of chemical compounds and their activities against biological assays are available on various platforms. As a consequence, there is a growing need for innovation in computer technology that can efficiently and accurately analyze ever-increasing amounts of physicochemical and biological data [5].

In the last years, quantum computing [6–8] has attracted much attention because it is one of the most promising quantum technologies that could radically transform science and many areas of industry. Although large-scale, fault-tolerant quantum computers have not yet been invented, noisy intermediate-scale quantum (NISQ) computers [6] have been applied to various areas of science and technology: chemistry [9–13], optimization [14–16], and finance [17, 18], to name but a few. A promising scheme for practical applications on NISQ devices is the hybrid quantum–classical algorithm [6, 8], in which computational tasks are deliberately divided into quantum and classical resources using a parameterized approach. Two important classes of such quantum algorithm are the variational quantum eigensolver for quantum simulation [19–21] and the quantum approximate optimization algorithm for combinatorial optimization [22–25].

More recently, quantum machine learning (QML) [27–31] is a rapidly growing research field that combines near-term quantum algorithms and machine learning techniques. In particular, parameterized quantum circuits (PQCs) have been considered as machine learning models with high expressive power within the hybrid quantum–classical framework [32, 33]. PQCs are typically composed of fixed quantum gates (e.g., qubit rotations and entangling gates) in a shallow circuit layout, with variable parameters optimized in a classical feedback loop. So far, QML has been successfully applied to both discriminative [34–38] and generative [39, 40] models. Examples of these include binary classification problems for image recognition [35], kernel methods for support vector machine [41, 42], and unsupervised machine learning in finance [43]. To our knowledge, however, the application of QML to regression tasks has not been fully investigated in the literature. It remains unclear what kinds of quantum states should be used in order to generate the feature map with high expressibility that is suited for real-world data sets.

To explore the possibility of near-term quantum applications to regression tasks, here we apply the QML method to quantitative structure–toxicity relationship (QSTR) models for predicting the toxicity of 221 phenols. While there are a variety of QSAR/QSTR models (e.g., 3D-QSAR [1, 4]), as a first step we employ QSAR/QSTR models including molecular descriptors such as hydrophobicity, acidity constant, and frontier orbital energies. There have been quantum computations in biochemical and pharmaceutical areas, such as protein folding [44–46], molecular similarity [47], and biological data [48]; yet, there has been no study on quantum application to QSAR modeling, albeit an important part of ligand-based computer-aided drug design.

The remainder of the paper is organized as follows. In Methods, we briefly review PQC-based machine learning and then describe our QML models in full detail. The information about the data set used for the QSTR modeling is also provided. In Results and Discussion, we present the results of our QSAR models and numerically investigate how different encodings, variational circuit architectures, and redundant encodings using multiple qubits affect the performance of the QML models. In addition, we compare the performance of our best QML models with those obtained by conventional chemometrics methods and comment on several perspectives on QML. Then, we summarize our conclusions.

## 2. Methods
### 2.1. Parameterized quantum circuits

In recent years, PQCs have been regarded as machine learning models with high expressive power within the framework of the hybrid quantum–classical approach. PQCs are usually composed of one-qubit rotations and two-qubit entangling operations in a shallow circuit layout, with parameters optimized in a feedback loop. A recent review on PQCs can be found in the literature [33]. Combining near-term quantum algorithms and machine learning, QML using the framework of PQCs is sometimes referred to as *quantum circuit learning* (QCL) [32]. So far, QML has been applied to both discriminative and generative tasks [34–43]; on the other hand, the application of QML to regression tasks has not been thoroughly investigated.

From the viewpoint of the machine learning architecture, PQCs consist of three components: the encoder circuit, the variational circuit, and the measurement for the estimation of the loss function. First, an encoder circuit loads classical $d$-dimensional data $\boldsymbol{x} = (x_1, x_2, \ldots, x_d)^\mathrm{T} \in \mathbb{R}^d$ into a higher-dimensional feature map $U_{\Phi(x)}$ in the Hilbert space, which produces a quantum state $U_{\Phi(x)}|0\rangle^{\otimes n}$, with $n$ being the number of qubits. The number of qubits $n$ can be set to the dimension of input data $d$ (other situations are also considered in the present work). Such approach may be less efficient in terms of the number of qubits but is efficient in terms of circuit depth. Second, a variational circuit $U(\boldsymbol{\theta})$ acts on the quantum state prepared by the encoder circuit, in order to explore the quantum-enhanced feature space using trainable parameters $\boldsymbol{\theta}$, leading to the parameterized quantum state $U(\boldsymbol{\theta})U_{\Phi(\boldsymbol{x})}|0\rangle^{\otimes n}$. Third, the loss function can be estimated from the expectation value by measurements. In the following subsections, we will closely look at each step of our QML models.

## 2.2. Encoder circuit

Data representation is essential for the success of machine learning models. In QML, loading classical data as a quantum state is an important and challenging task; in fact, the choice of encoding in PQCs is analogous to selecting a feature map in kernel-based machine learning techniques [33, 42]. Several methods for encoding input data into qubits have been proposed: amplitude encoding [33, 42], angle encoding [34, 49], a random linear map [35], and data re-uploading [38]. However, it is not a priori obvious what kind of encoding is suitable for our particular application. With this in mind, we first employ three methods of loading classical data into a quantum state (note that we can pre-process input data by means of normalization).

A first encoding is the one proposed by Mitarai et al. [32]:

$$\mathcal{U}_{\mathrm{M}} = \prod_{q=1}^{n} R_q^Z(\cos^{-1} x_q^2) R_q^Y(\sin^{-1} x_q).$$

(1)

This approach was originally motivated by expanding the density operator of a quantum state in terms of a set of Pauli operators [32]. A second encoding we consider is an angle encoding [34, 49] and the corresponding unitary operator $\mathcal{U}_{\mathrm{A1}}$ can be defined by

$$\mathcal{U}_{\mathrm{A1}} = \prod_{q=1}^{n} R_q^Y(x_q).$$

(2)

This scheme is sometimes referred to as *qubit encoding* [50]. The encoding can be viewed as the product of local kernels, where each component of the input vector is encoded into a local feature map; it has the same structure as a product quantum state that is unentangled [33]. This kind of encoding, though seemingly simple, has been applied for tree tensor network classifiers in QML [34]. A third encoding is related to the second one and uses a couple of single-qubit rotations. The corresponding unitary operator $\mathcal{U}_{\mathrm{A2}}$ can be expressed as

$$\mathcal{U}_{A2} = \prod_{q=1}^{n} R_q^Z(x_q) R_q^Y(x_q).$$

(3)

This encoding loads each component of the input vector into two angles in the Bloch sphere, generating a certain redundancy in encoding and hence the possible modification in the feature map.

In addition to investigating different ways of encoding, we explore the possibility that entanglement might extend the flexibility in data representation. In fact, the previous studies suggest that entangling gates play essential roles in quantum generative models [40, 51] and in expressibility for PQCs [52]; in particular, repeated circuit layers with entangling controlled NOT (CNOT) gates provide high expressive power [51, 52]. In this work, we propose an encoder circuit containing entangler blocks in data representation. Such encoding circuit can be expressed as multiple layers of single-qubit rotations followed by two-qubit entangling gates:

$$V_{\Phi(x)} = \prod_{k=1}^{K} E_{\text{ent}}^{k} \mathcal{U}_{\phi_k(x)}.$$

(4)

Here, the $k$th layer of the operations comprises a product of two operations: (i) the unitary operator $\mathcal{U}_{\phi_k(x)}$ that is any reasonable encoder circuit loading classical input data $x$ and (ii) the two-qubit entangling operation $E_{\text{ent}}^{k}$ that is typically composed of CNOT or controlled Z (CZ) gates (which are hereafter denoted as $E_{\text{CNOT}}$ and $E_{\text{CZ}}$, respectively). In the following, the encoding described in equation (4) is referred to as *entangler-enhanced encoding*. We could expect that such encoding might expand the representation ability in the feature map, owing to quantum entanglement. From the viewpoint of quantum physics, the above encoding can be interpreted as a concatenated tensor network and this family of quantum circuits can describe a high dimensional tensor network in an efficient way [53]. In the present study, we consider the following encoding composed of two layers:

$$V_{\Phi(x)} = E_{\text{ent}} \mathcal{U}_2 E_{\text{ent}} \mathcal{U}_1,$$

(5)

where the unitary operators $\mathcal{U}_1$ and $\mathcal{U}_2$ can be any of the three encodings mentioned earlier. Our approach can be viewed as an extension of the previous QCL [32], where the feature map is represented by the product state. To investigate the performance of our entangler-enhanced encoders, we considered 10 combinations for $\mathcal{U}_1$, $\mathcal{U}_2$, and $E_{\text{ent}}$, which are summarized in table 1.

Another approach to increase the flexibility of the feature space is to use $p$ copies of quantum states (i.e., the $p$-times product) at the outset, which means that each component of the input data is encoded into multiple qubits [32, 42]. While this scheme requires additional quantum resources, it generates higher-order terms in the feature map, which is likely to give rise to more expressive power and a richer class of functions. A recent study indicates that such input redundancy is necessary for the task of data fitting and that it grows at least logarithmically in the complexity of the functions [54]. For each encoding in table 1, we thus consider feature maps using two and three copies of the quantum states in encoding.

**Table 1.** Encoder circuits investigated in the present work ($E_{\text{ent}}\mathcal{U}_2 E_{\text{ent}}\mathcal{U}_1$) and the corresponding ID. Note that the first three encodings are conventional encoders (i.e., $E_{\text{ent}}$ and $\mathcal{U}_2$ are replaced by the identity operator) whereas the remaining 10 encoders contain entangler blocks $E_{\text{ent}}$. For the definitions of the unitary operations, see the text.

| Encoder Circuit ID | $\mathcal{U}_1$ | $\mathcal{U}_2$ | $E_{\text{ent}}$ |
|---|---|---|---|
| M | $\mathcal{U}_M$ | — | — |
| A1 | $\mathcal{U}_{A1}$ | — | — |
| A2 | $\mathcal{U}_{A2}$ | — | — |
| M–M–CNOT | $\mathcal{U}_M$ | $\mathcal{U}_M$ | $E_{\text{CNOT}}$ |
| A1–A1–CNOT | $\mathcal{U}_{A1}$ | $\mathcal{U}_{A1}$ | $E_{\text{CNOT}}$ |
| A2–A2–CNOT | $\mathcal{U}_{A2}$ | $\mathcal{U}_{A2}$ | $E_{\text{CNOT}}$ |
| M–A1–CNOT | $\mathcal{U}_M$ | $\mathcal{U}_{A1}$ | $E_{\text{CNOT}}$ |
| M–A2–CNOT | $\mathcal{U}_M$ | $\mathcal{U}_{A2}$ | $E_{\text{CNOT}}$ |
| M–M–CZ | $\mathcal{U}_M$ | $\mathcal{U}_M$ | $E_{\text{CZ}}$ |
| A1–A1–CZ | $\mathcal{U}_{A1}$ | $\mathcal{U}_{A1}$ | $E_{\text{CZ}}$ |
| A2–A2–CZ | $\mathcal{U}_{A2}$ | $\mathcal{U}_{A2}$ | $E_{\text{CZ}}$ |
| M–A1–CZ | $\mathcal{U}_M$ | $\mathcal{U}_{A1}$ | $E_{\text{CZ}}$ |
| M–A2–CZ | $\mathcal{U}_M$ | $\mathcal{U}_{A2}$ | $E_{\text{CZ}}$ |

## 2.3. Variational circuit

The essential role of the variational circuit $U(\boldsymbol{\theta})$ is to explore efficiently the quantum-enhanced feature space generated by PQCs. The variational circuit originally reported in the literature is based on the time evolution of Ising Hamiltonian [32]; it uses the Trotter decomposition method, which requires an additional computational cost. Another disadvantage of the method is that it is rather memory-intensive when performing quantum simulator on classical processors.

To circumvent the limitations, we employed quantum circuits inspired by the strategy of the hardware-heuristic ansatz [13], which was originally motivated by the limitations of existing NISQ devices in fidelity and connectivity. On the basis of the architecture of PQCs [13, 52], here we propose that the variational circuit be constructed by $L$ layers of the unit circuit consisting of single-qubit rotations $U_\ell(\boldsymbol{\theta}_\ell)$ and two-qubit entangler blocks $E_{\text{ent}}^\ell$ comprising CNOT or CZ gates:

$$U(\boldsymbol{\theta}) = \prod_{\ell=1}^{L} U_\ell(\boldsymbol{\theta}_\ell) E_{\text{ent}}^\ell.$$

(6)

From a physical standpoint, such quantum circuit can be interpreted as a concatenated tensor network, which can be used for an efficient description of time-evolved quantum states [53]. In this work, we investigated the performance of the three variational circuits (see figure 1): one was the variational circuit based on the time evolution of Ising Hamiltonian and the other two were the modified variational circuits based on the hardware-heuristic approach (the total number of two-qubit gates is $nL$). In both approaches, the total number of trainable parameters is $3nL$.

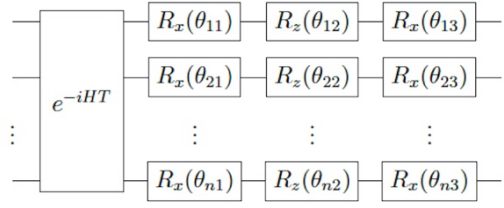

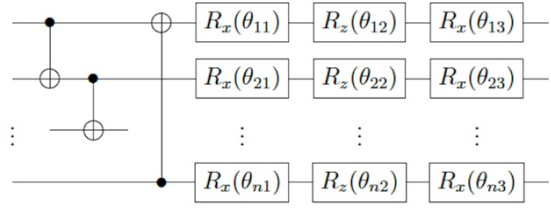

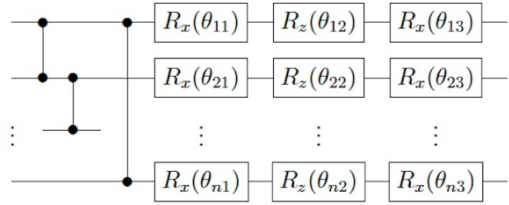

**Figure 1.** Circuit unit (template) used for the variational circuit $U(\boldsymbol{\theta})$: (a) circuit unit exploiting on the time evolution of Hamiltonian, (b) circuit unit containing a CNOT block $E_{\mathrm{CNOT}} = \prod_{q=1}^{n} \mathrm{CNOT}_{q,(q+1) \bmod n}$, and (c) circuit unit containing a CZ block $E_{\mathrm{CZ}} = \prod_{q=1}^{n} \mathrm{CZ}_{q,(q+1) \bmod n}$. Note that each circuit unit can be repeated $L$ times in the variational circuit. The total number of trainable parameters in $U(\boldsymbol{\theta})$ is thus $3nL$; the total number of two-qubit entangling gates in $U(\boldsymbol{\theta})$ is $nL$ in the cases of (b) and (c).

## 2.4. Measurements and supervised learning

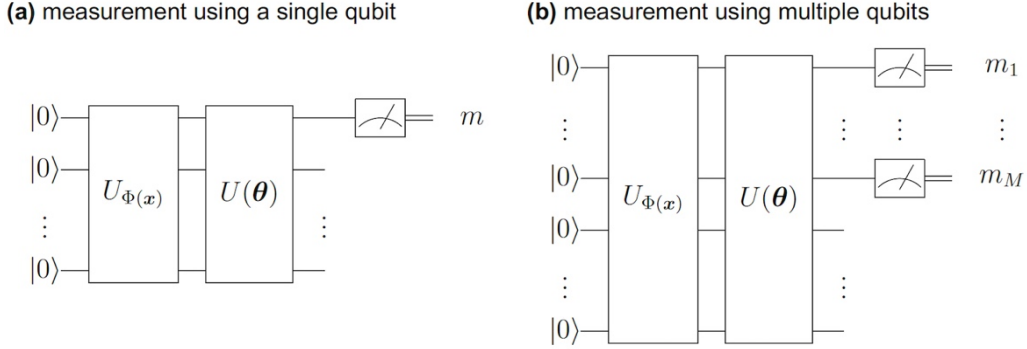

**Figure 2.** Measurement of (a) an expectation value from a single qubit or (b) a set of expectation values from multiple qubits.

In QML, the measurement of quantum states extracts the information that can be used for supervised learning. For instance, a QML architecture can measure an expectation value by acting a Pauli $Z$ operator on a single qubit (figure 2(a)). This expectation value can be used for the evaluation of the loss function. Since the information is reduced to only one qubit and then extracted by the measurement, this approach may be considered as *pure* QML (unless otherwise mentioned, this scheme was employed in this work). For the values $\mathbf{y} = (y^{(1)}, y^{(2)}, \ldots, y^{(N)})^{\mathrm{T}} \in \mathbb{R}^N$ (where $N$ is the number of data samples) and the expectation values $\{m^{(i)}\}$, the loss function $\mathcal{L}$ can be given by

$$\mathcal{L}(\{y^{(i)}\}, \{m^{(i)}\}) = \frac{1}{N} \sum_{i=1}^{N} (y^{(i)} - m^{(i)})^2. \tag{7}$$

Note that $\{y^{(i)}\}$ are normalized between 0 and 1. Another approach is to use a set of expectation values from multiple qubits [33] for the evaluation of the loss function. This scheme can be viewed as hybrid quantum–classical machine learning and such quantum circuits are also considered in this work (figure 2(b)). By *hybrid* QML, we mean that the measurements of quantum states generated by PQCs are used as input for an additional classical machine learning model. Note that the hybrid quantum-classical algorithm is a more general framework for NISQ applications. As a first step, we simply use the expectation values as input for a multiple linear model. For a set of expectation values from $M$ qubits for $i$th data, $\{m_q^{(i)}\}_{q=1}^{M}$, the predicted value $\hat{y}^{(i)}$ can be expressed as

$$\hat{y}^{(i)} = \sum_{q=1}^{M} \beta_q m_q^{(i)},$$

(8)

with the coefficient vector $\boldsymbol{\beta} = (\beta_1, \beta_2, \ldots, \beta_M)^{\mathrm{T}} \in \mathbb{R}^M$. In this linear model, the loss function $\mathcal{L}$ can be estimated by

$$\mathcal{L}\left(\{y^{(i)}\}, \{m_q^{(i)}\}_{q=1}^{M}\right) = \frac{1}{N}\sum_{i=1}^{N}\left(y^{(i)} - \sum_{q=1}^{M}\beta_q m_q^{(i)}\right)^2 = \frac{1}{N}\|\boldsymbol{y} - \mathcal{M}^{\mathrm{T}}\boldsymbol{\beta}\|_2^2,$$

(9)

with $\{\mathcal{M}_{qi}\} := \{m_q^{(i)}\} \in \mathbb{R}^{M \times N}$ and the optimized $\boldsymbol{\beta}^* = (\mathcal{M}\mathcal{M}^{\mathrm{T}})^{-1}\mathcal{M}\boldsymbol{y}$.

In our regression tasks, we used a standard approach that minimizes the loss function with respect to trainable parameters $\boldsymbol{\theta}$. In the present work, a regularization term was not included, since overfitting would be effectively avoided owing to the inherent constraints arising from the unitary conditions [32]. In minimizing the loss function, we used the Nelder–Mead method [55], which is a gradient-free algorithm. In our QML models, the scaling factor for observable quantities from the measurements $f_{\langle Z \rangle}$ is a hyperparameter. Thus, we systematically varied the hyperparameter in our simulations. Values for the scaling factor for the expectation value $\langle Z \rangle$ were chosen between 1.0 and 10.0, depending on models (the values for the scaling factor were tuned by the grid search scheme). We used mean squared error (MSE) to evaluate the error of a prediction. The coefficients of determination ($R^2$) were calculated for evaluating the performance of our QSTR models.

## 2.5. Implementation

We implemented our QML models using `Qulacs` [56], a `Python/C++` library for quantum circuit simulation. The time evolution gate of Ising Hamiltonian needed for the original QCL model was implemented using `NumPy` [57] and `SciPy` [58] libraries. The Nelder–Mead optimization of Pauli rotation angles was implemented using `scipy.optimize` module in `SciPy` library. The *k*-fold cross-validation was implemented using `KFold` module in `scikit-learn` [59] library. Pre- and post-processing of the data set was implemented using `pandas` [60] library in combination with `NumPy` and `SciPy` libraries.

2.6. Data set

In our QSTR models, we used a data set of 221 phenols, for which toxicity data to the ciliate *Tetrahymena pyriformis* in terms of $\log(1/\text{IC}_{50})$ are available [61]. We used the following molecular descriptors: hydrophobicity ($\log K_{\text{ow}}$), acidity constant ($pK_a$), frontier orbital energies ($E_{\text{HOMO}}$ and $E_{\text{LUMO}}$), and hydrogen bond donor/acceptor counts ($N_{\text{hdon}}$). The data set has been used for evaluating the performance and predictive abilities of standard chemometrics methods [61–63]: multiple linear regression (MLR), support vector machine, and radial basis function neural networks (RBF-NNs). To compare our QML models with conventional chemometrics methods, we trained MLR and RBF-NN models on the same data set (we did not train an SVM model because the performance of SVM is comparable to that of RBF-NN in QSAR studies [62]). Following the previous QSTR study [63], we used the hold-out validation for our QSTR models; more specifically, we used 180 compounds for a training set and 41 for a validation set. Note that the data splitting we used for the hold-out validation was exactly the same as in the previous work [63], in which the Kennard–Stone algorithm [64] was employed for generating the data splitting in order to make all the validation data fall inside the training data. Such data splitting is useful because the data set in the QSTR study is somewhat widely distributed and contains certain outliers. We also performed 5-fold cross-validation on the entire data set that had been randomly sorted.

2.7. Simulation details

All of the simulations for QML, MLR, and RBF-NN models were performed on a classical computational platform, powered by Intel Xeon Gold 6154 processors with 192 GB memory. All the simulations except for the QML models with 15 qubits were performed using a single CPU core; and the QML simulations with 15 qubits were performed by `OpenMP` parallel jobs using 9 CPU cores.

## 3. Results and discussion
### 3.1. Encoder circuit

Table 2. Coefficients of determination for the training and the validation sets ($R^2_{\text{train}}$ and $R^2_{\text{val}}$) using 13 different encoder circuits with 5 qubits, the optimized number of layers in the variational circuit $L$ ($3 \leq L \leq 12$) and the scaling factor $f_{\langle Z \rangle}$ for the expectation value $\langle Z \rangle$.

| Encoder Circuit ID | $R^2_{\text{train}}$ | $R^2_{\text{val}}$ | $L$ | $f_{\langle Z \rangle}$ |
|---|---|---|---|---|
| M | 0.656 | 0.810 | 10 | 10.0 |
| M–M–CZ | 0.682 | 0.843 | 9 | 10.0 |
| M–A1–CZ | 0.776 | 0.855 | 11 | 8.0 |
| M–A2–CZ | 0.820 | 0.821 | 12 | 2.0 |
| M–M–CNOT | 0.740 | 0.836 | 11 | 8.0 |
| M–A1–CNOT | 0.784 | 0.833 | 7 | 2.0 |
| M–A2–CNOT | 0.819 | **0.849** | 6 | 2.0 |
| A1 | 0.777 | 0.824 | 7 | 2.0 |
| A2 | 0.735 | 0.817 | 8 | 4.0 |
| A1–A1–CZ | 0.822 | 0.848 | 11 | 2.0 |
| A2–A2–CZ | 0.774 | 0.805 | 8 | 4.0 |
| A1–A1–CNOT | 0.808 | 0.828 | 12 | 1.0 |
| A2–A2–CNOT | **0.842** | 0.844 | 11 | 1.0 |

**Table 3.** Coefficients of determination for the training and the validation sets ($R^2_{\text{train}}$ and $R^2_{\text{val}}$) using 13 different encoder circuits with 10 qubits (two copies of the quantum states), the optimized number of layers in the variational circuit $L$ ($3 \leq L \leq 12$) and the scaling factor $f_{\langle Z \rangle}$ for the expectation value $\langle Z \rangle$.

| Encoder Circuit ID | $R^2_{\text{train}}$ | $R^2_{\text{val}}$ | $L$ | $f_{\langle Z \rangle}$ |
|---|---|---|---|---|
| M | 0.773 | 0.807 | 10 | 8.0 |
| M–M–CZ | 0.806 | 0.830 | 10 | 10.0 |
| M–A1–CZ | 0.846 | 0.842 | 8 | 6.0 |
| M–A2–CZ | 0.881 | 0.839 | 12 | 8.0 |
| M-M-CNOT | 0.816 | 0.827 | 12 | 10.0 |
| M–A1–CNOT | 0.857 | 0.851 | 9 | 10.0 |
| M–A2–CNOT | 0.873 | 0.852 | 9 | 6.0 |
| A1 | 0.828 | 0.851 | 5 | 10.0 |
| A2 | 0.875 | 0.843 | 12 | 10.0 |
| A1–A1–CZ | 0.893 | 0.853 | 9 | 10.0 |
| A2–A2–CZ | 0.823 | 0.842 | 6 | 8.0 |
| A1–A1–CNOT | 0.881 | 0.862 | 9 | 8.0 |
| A2–A2–CNOT | **0.906** | **0.869** | 10 | 10.0 |

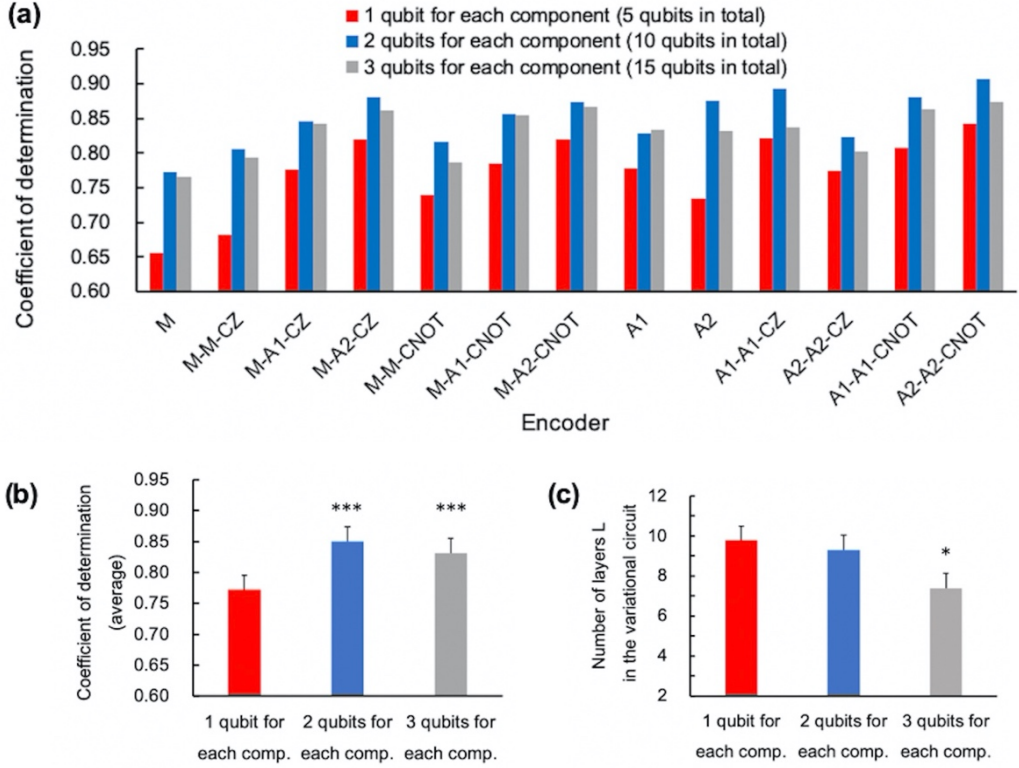

**Figure 3.** Performance for the training set using 13 different encoder circuits in the cases of 5 (red), 10 (blue), and 15 (grey) qubits. (a) Coefficients of determination for the training set ($R^2_{\text{train}}$) using 13 different encoder circuits (for the definitions of the encoder circuits, see table 1). (b) Average $R^2_{\text{train}}$ values for 5, 10, and 15 qubits. (c) Average number of layers $L$ in the variational circuit for 5, 10, and 15 qubits. In (b) and (c), error bars indicate the standard error and asterisks indicate statistical significance in comparison with the case of 5 qubits (paired *t*-test; *=$p < 0.05$; ***=$p < 0.0005$).

To begin with, we compared the performance of the three conventional encodings with 5 qubits (table 2). As to the performance of $R^2_{\text{train}}$, the A1 and A2 encoders (0.777 and 0.735) performed better by 15% than the M encoder (0.658). The results indicate that the angle encodings provide more flexibility in data encoding owing to its simplicity and high nonlinearity. To improve the performance of our QML models, we then explored the possibility that entanglement might extend expressive power in data representation. It has been shown that entangling gates play essential role in quantum generative models [40, 51] and in expressibility for PQCs [52].

We employed the encoder circuits having CNOT or CZ gates (table 2). As to the performance of $R^2_{\text{train}}$, our entangler-enhanced encodings containing $\mathcal{U}_M$ performed better by 15% than the original $\mathcal{U}_M$ unitary. In the case of the angle encodings, the encoders containing entangling gates outperformed those without entanglement by 7%. In particular, the A2–A2–CNOT encoder provided the best performance (0.842), followed by that obtained by the A1–A1–CZ encoder (0.822). This result is consistent with the previous studies on PQCs, in which repeated circuit layers with entangling gates provide high expressive power [52]. Our results indicate that the feature map using the product state was inadequate for our application in terms of expressibility and that the entangler-enhanced encodings provided more expressive power in data representation with the aid of quantum entanglement. This implies that quantum correlation could be advantageous for the feature map representation of classical input data.

To comprehend the roles of the redundancy in encoding associated with higher-dimensional local feature maps, we then increased the number of qubits in our QML models. In this scheme, each component of the input data is encoded into multiple qubits. Recently, Vidal and Theis investigated whether the redundancy in PQCs is useful for the task of data fitting [54]; and their study indicates that lower bounds of the redundancy are logarithmic in terms of the complexity of the functions. Since five molecular descriptors were contained in our QSTR models, we used 10 and 15 qubits, which corresponds to two and three qubits per input data, respectively.

The use of 10 qubits (two copies of the quantum states) led to a 10% increase in $R^2_{\text{train}}$ in comparison with the case of 5 qubits (figures 3(a) and (b)). The paired $t$-test also suggested that the difference in $R^2_{\text{train}}$ between 5- and 10-qubit cases was statistically significant. The results indicate that, in our QML models with 10 qubits, higher dimensionality was effectively taken into account owing to the redundancy in multiple-qubit encoding. By encoding each component of the input data into the higher-dimensional local feature map, the encoder is composed of a more complete basis of functions and can respond to smaller changes in the input data [50]. In line with the results with 5 qubits, the A2–A2–CNOT encoder provided the best performance in $R^2_{\text{train}}$ (0.906). (table 3). This confirms that our entangler-enhanced encodings provided more flexibility in data representation. The encoders containing CNOT gates had the tendency to perform better than those containing CZ gates. This expressive power might be related to the fact that increasing CNOT gates in multilayer PQCs leads to an increase in the bond dimensions in tensor networks [51].

The average values for $R^2_{\text{train}}$ were 0.772, 0.851, and 0.832 for 5, 10 and 15 qubits, respectively (see also figure 3(b) and table A1). According to the paired $t$-test, the difference in $R^2_{\text{train}}$ between 5- and 15-qubit cases was also statistically significant (figure 3(b)). The average numbers of layers $L$ in $U(\boldsymbol{\theta})$ were 9.8, 9.3, and 7.4 for 5, 10, 15 qubits, respectively, indicating that the number of layers $L$ was decreased with respect to the number of qubits (figure 3(c)). These numerical results indicate that, in our QML models, encoding each component of the input vector into multiple qubits appeared to be helpful for improving the performance. It is possible that the performance could experience some sort of saturation in terms of the number of qubits. A computational aspect is that increasing the number of qubits causes an increase in the number of trainable parameters in $U(\boldsymbol{\theta})$, which could result in the slower convergence in minimizing the cost functions. In this particular application, our QML models using the entangler-enhanced encoding with 10 qubits appeared to give good results. Our results agree with the previous study on quantum classifiers using data re-uploading [38], in which using more qubits and entanglement increases the classification success rate and reduces the number of layers required.

3.2. Variational circuit

To understand how the architecture of the variational circuit affects the performance and the computational cost, we used the three variational circuits while using the same encoding circuit (the M encoder). The first variational circuit was the one based on Ising Hamiltonian, which was previously proposed [32]. The second and third circuits were CNOT-based and CZ-based variational circuits, respectively. The latter two circuits are motivated by the strategy of hardware heuristic ansatz in order to circumvent the limitations of quantum hardware; the two circuits can avoid an additional computational cost generated by the Trotter decomposition.

According to our numerical tests on simple regression tasks, we found that CNOT-based variational circuit provided a similar performance compared with the variational circuit based on Ising Hamiltonian, whereas CZ-based variational circuit gave an inferior performance (table B1). The results indicate that repeated circuit layers with entangling CNOT gates provide high expressive power, in line with the previous studies, where CNOT gates play important roles in expressibility of PQCs [51, 52]. Therefore, we employed the variational circuit containing entangling CNOT gates, unless otherwise mentioned. In addition, we observed the substantial computational speedup by using the variational circuit containing entangling gates, compared with the original variational circuit based on Ising Hamiltonian. A major disadvantage of the latter is that the computational cost and the memory requirement for the calculation of the Trotter operator matrix grows exponentially with respect to the number of qubits, for quantum simulator on classical processors (table B2). For that reason, we recommend the use of the hardware-heuristic variational circuits.

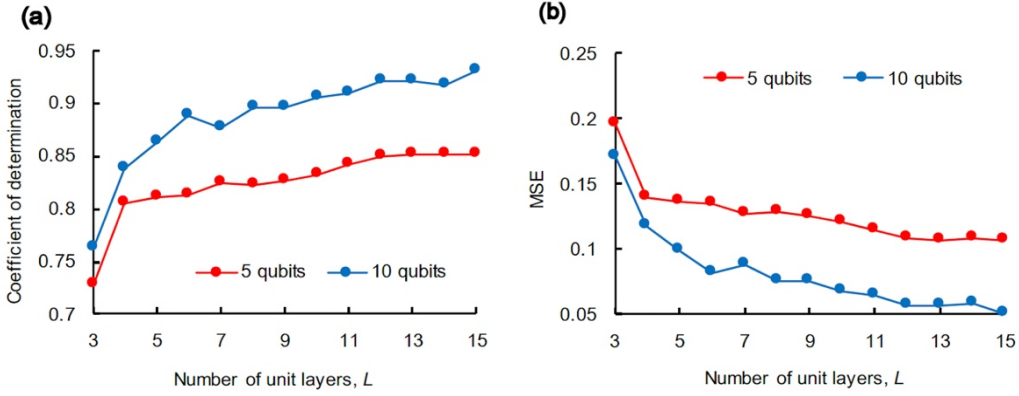

**Figure 4.** Coefficients of determination $R^2_{\text{train}}$ (a) and MSE (b) for the training set as a function of the number of unit layers $L$ in $U(\boldsymbol{\theta})$. The results were obtained using the A2–A2–CNOT encoder with 5 (red) and 10 (blue) qubits. For the definition of the A2–A2–CNOT encoder, see table 1.

Furthermore, we checked the effects of the number of unit layers $L$ on the performance of our QML models. According to our simulations ($3 \leq L \leq 15$), adding unit layers normally provided good results; a typical example of this tendency can be found in figure 4(a), in which $R^2_{\text{train}}$ obtained using the A2–A2–CNOT encoder is gradually improved as a function of $L$ (as to the case of 5 qubits, the performance appears to be saturated for $L \geq 12$). This is consistent with the decrease in MSE for the training set by increasing $L$ (figure 4(b)). The results imply an improved efficiency in exploring the solution space by adding circuit unit layers, in agreement with the previous studies on PQCs [51, 52]. We also found that the optimized numbers of layers in our QML models was significantly dependent on the choice of the PQC architecture and of the encoding (see also tables 2 and 3). A similar tendency has been reported in the previous work on the expressibility and entangling capability of PQCs, in which the rates of change in expressibility with respect to the number of layers tend to vary from circuit to circuit [52].

## 3.3. Final QML model

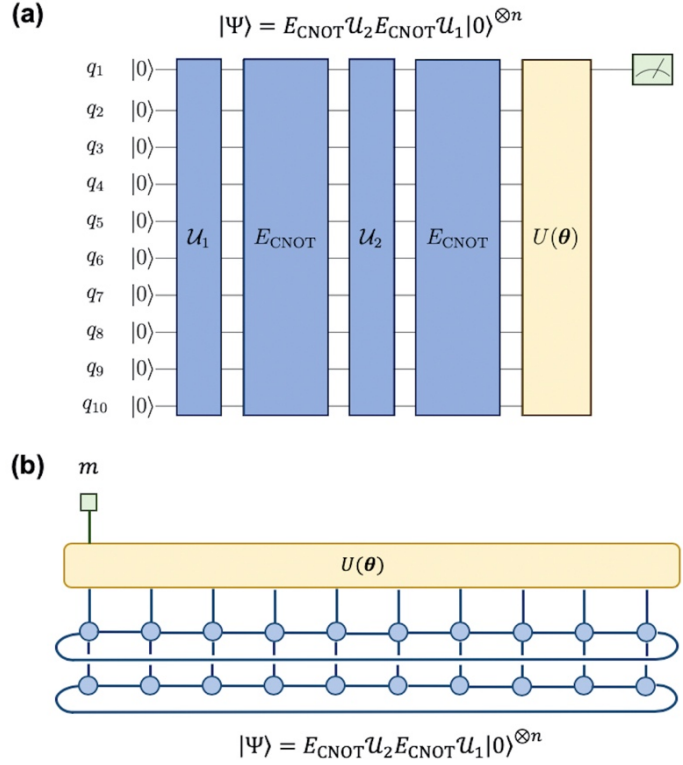

**Figure 5.** Quantum circuit (a) and the graphical tensor network representation (b) for our QML model suitable for our QSTR study. The feature map is given by $|\Psi\rangle = E_{\text{CNOT}}\mathcal{U}_2 E_{\text{CNOT}}\mathcal{U}_1|0\rangle^{\otimes n}$ (in our best QML model, $\mathcal{U}_1 = \mathcal{U}_2 = \mathcal{U}_{\text{A2}}$). From a physical standpoint, such quantum circuit can be interpreted as a 2D tensor network, in which the entangler block can be interpreted as the periodic boundary condition. Each component of the input data is encoded into two qubits (i.e., $n = 10$ qubits) in order to increase dimensionality in the feature map. The variational circuit is given by $U(\boldsymbol{\theta}) = \prod_{\ell=1}^{L} U_\ell(\boldsymbol{\theta}_\ell) E_{\text{CNOT}}^{\ell}$, which is a multilayer PQC (for the inner details of the variational circuit $U(\boldsymbol{\theta})$, see figure 1(b)). The measurement can be done by either a single qubit or multiple qubits (see also figure 2). Note that only a single-qubit measurement is depicted.

Considering the results presented in the previous subsections, we obtained our final QML model suitable for our particular application (depicted in figure 5). Our final model can be described as follows. The quantum circuit for data representation is given by the entangler-enhanced encoder $E_{\text{CNOT}}\mathcal{U}_2 E_{\text{CNOT}}\mathcal{U}_1$; in our best QML model, $\mathcal{U}_1 = \mathcal{U}_2 = \mathcal{U}_{\text{A2}}$ (i.e., the A2–A2–CNOT encoder) (figure 5(a)). Hence, the feature map can be given by $|\Psi\rangle =$ $E_{\text{CNOT}}\mathcal{U}_{\text{A2}}E_{\text{CNOT}}\mathcal{U}_{\text{A2}}|0\rangle^{\otimes n}$. This kind of encoder can be viewed as a 2D tensor network, in which the entangler block can be interpreted as the periodic boundary condition (figure 5(b)). Each component of the input data is encoded into two qubits, meaning that the feature map with higher dimensionality can be taken into account; consequently, 10 qubits are used for encoding because five molecular descriptors are contained in our QSTR model. In this way, our final QML model allows multiple data encoding, which is consistent with the idea of input redundancy [54] or data re-uploading [38]. The variational circuit $U(\boldsymbol{\theta})$ is given by a multilayer PQC: $U(\boldsymbol{\theta}) = \prod_{\ell=1}^{L} U_\ell(\boldsymbol{\theta}_\ell) E_{\text{CNOT}}^{\ell}$.

### 3.4. Measurements and the hybrid approach

We compared the performance between pure and hybrid QML models for the A2–A2–CNOT encoder with 10 qubits (figure C1). Overall, the values for $R^2_{\text{train}}$ were improved by about 2% when using the hybrid QML approach, in which the expectation values from $M$ qubits were fed into the evaluation of the loss function. However, increasing the number of qubits for the measurements $M$ did not necessarily lead to incremental improvements in the $R^2_{\text{train}}$ performance. Rather, we found that the number of unit layers $L$ in $U(\boldsymbol{\theta})$ had an overall impact on the $R^2_{\text{train}}$ performance. Also, there were quite a few cases where the performance on the validation set were not improved, compared with those obtained by pure QML models (this topic will be discussed in the next subsection). On the other hand, we found that the QML model with $M = 4$ and $L = 10$ provided the best performance for $R^2_{\text{val}}$ (0.886). Further improvement for the post-processing on the classical part (e.g., classical neural networks) may be necessary.

## 3.5. Performance comparison with conventional chemometrics

Table 4. Performance comparison of our QML models with those obtained from conventional chemometrics methods (the coefficients of determination, MSE, and root mean square (RMS) for the training and the validation sets).

| Methods | $R^2_{train}$ | $R^2_{val}$ | $MSE_{train}$ | $MSE_{val}$ | $RMS_{train}$ | $RMS_{val}$ |
|---|---|---|---|---|---|---|
| QML(A2–A2–CNOT-10q-m4) [a] | 0.913 | **0.886** | 0.062 | **0.046** | 0.250 | **0.214** |
| QML(A2–A2–CNOT-10q) [b] | 0.906 | 0.869 | 0.067 | 0.052 | 0.260 | 0.229 |
| QML (original-5q) [c] | 0.644 | 0.825 | 0.256 | 0.070 | 0.506 | 0.264 |
| RBF-NN (ref. 63) | **0.942** | 0.882 | **0.041** | 0.058 | **0.204** | 0.240 |
| RBF-NN [d] | 0.928 | 0.819 | 0.052 | 0.072 | 0.227 | 0.269 |
| MLR (ref. 63) | 0.602 | 0.786 | 0.286 | 0.102 | 0.535 | 0.320 |
| MLR [d] | 0.609 | 0.740 | 0.281 | 0.104 | 0.530 | 0.322 |

[a] Obtained using the A2–A2–CNOT encoder (10 qubits) combined with the hardware-heuristic variational circuit ($M = 4$; $L = 10$). [b] Obtained using the A2–A2–CNOT encoder (10 qubits) combined with the hardware-heuristic variational circuit ($L = 10$). [c] Obtained using the original encoder (5 qubits) combined with the variational circuit based on Ising Hamiltonian. [d] Calculated in the present work.

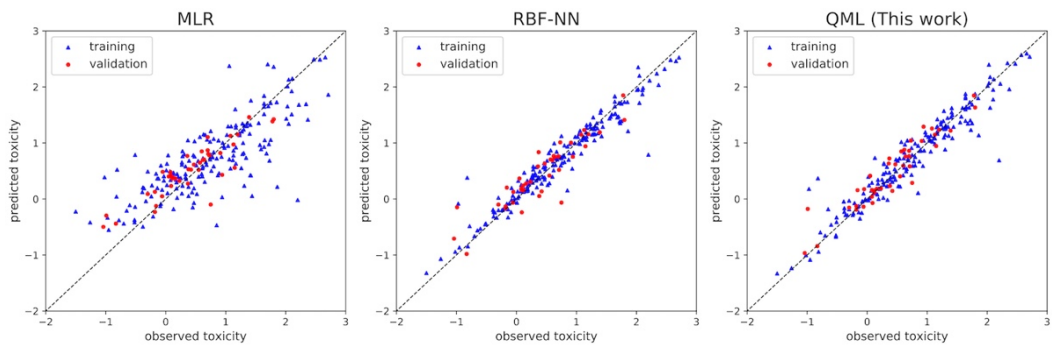

Figure 6. Plots for the observed vs. predicted toxicity obtained from the MLR, RBF-NN, and QML models (blue triangle: training set; red circle: validation set). (Left) MLR model ($R^2_{train}$: 0.609; $R^2_{val}$: 0.740). (Center) RBF-NN model ($R^2_{train}$: 0.928; $R^2_{val}$: 0.819). (Right) QML model ($R^2_{train}$: 0.906; $R^2_{val}$: 0.869) using the A2–A2–CNOT encoder with 10 qubits.

Having developed our QML models for QSTR application, we now compare their performance with those obtained by conventional chemometrics methods, namely MLR and RBF-NN methods (table 4). In our MLR model, the values for $R^2_{\text{train}}$ and $R^2_{\text{val}}$ were 0.609 and 0.740, respectively, in agreement with those in the previous work [63] (0.602 and 0.786, respectively). The values for $R^2_{\text{train}}$ and $R^2_{\text{val}}$ in the QML model using the original scheme were 0.644 and 0.825, respectively; and the average values for $R^2_{\text{train}}$ and $R^2_{\text{val}}$ obtained by our improved QML models were 0.910 and 0.878, respectively (51% and 12% higher than the MLR counterparts, respectively). Thus, the QML models performed significantly better than the MLR models, suggesting that the QML models succeeded in nonlinear regression tasks.

It is also important to compare the performance of our QML models with those obtained from the RBF-NN models (table 4), because RBF networks are capable of universal approximation [65]. The values for $R^2_{\text{train}}$ and $R^2_{\text{val}}$ in the previous RBF-NN model [63] are 0.942 and 0.882, respectively; and those in our RBF-NN model were 0.928 and 0.819, respectively. Hence, the performance of our QML models was comparable to those obtained by the RBF-NN models. In fact, the plots for the observed vs. predicted toxicity obtained by the RBF-NN and the QML models are remarkably similar to each other; and there is also a similarity in the distributions of certain outliers (figure 6). The results indicate that our quantum-enhanced feature map generated by the PQCs was similar to the mapping obtained by the RBF network, which is capable of universal approximation. Furthermore, our hybrid QML model ($M = 4$ and $L = 10$) provided a slightly better $R^2_{\text{val}}$ value (0.886) (see also figure C2), compared with the RBF-NN counterpart (0.882). This interpretation can be supported by the smallest $\text{MSE}_{\text{val}}$ and $\text{RMS}_{\text{val}}$ values (0.046 and 0.214, respectively) obtained by the hybrid QML model (table 4). Our results imply a high expressive power of our QML models using multilayer PQCs.

In order to shed light into the generalization performance of the modeling schemes, we further conducted 5-fold cross-validation using the entire data set. Note that this validation scheme is different from the hold-out validation in table 4, in which the data splitting was obtained using the Kennard–Stone algorithm. We found that RBF-NN and hybrid QML models tended to experience overfitting: for the RBF-NN method, the values for $R^2_{\text{train}}$ and $R^2_{\text{val}}$ were 0.933 and 0.619, respectively; for the hybrid QML approach, the values for $R^2_{\text{train}}$ and $R^2_{\text{val}}$ were 0.932 and 0.479, respectively. On the other hand, pure QML models (using the single qubit measurement) appeared to avoid overfitting: the values for $R^2_{\text{train}}$ and $R^2_{\text{val}}$ were 0.876 and 0.694, respectively; this is probably because the unitary conditions innately acted as regularization [32]. Considering all the results in this work, the performance comparison can be summarized as follows: our best QML model ≈ RBF-NN models > MLR models. Our results thus imply that the QML method could be an alternative approach for nonlinear regression tasks.

3.6. Perspectives on QML

Let us comment on several perspectives on QML models. While definitive quantum advantage for machine learning has been controversial, we anticipate that there may be several merits for employing QML. First, we can directly manipulate the feature map in terms of quantum many-body states. If one could use complex, computationally intractable quantum states as feature maps while avoiding overfitting, then that could be an advantage. Second, once the architecture of PQCs is designed, it can train QML models in an efficient way, without the need for further tuning. In particular, the unitary conditions inherent to quantum circuits can act as built-in regularization, which may result in the avoidance of overfitted models and the improvement of generalization performance. In the case of RBF-NNs, on the contrary, centers of the RBFs, the number of hidden layer units, widths, and weights have to be determined carefully. Third, QML models using PQCs require much less number of trainable parameters [37, 51] and perhaps fewer hyperparameters, implying the possibility of efficient and unbiased machine learning using near-term quantum computing. Fourth, on numerical simulators, the interpretation of QML models could be possible by analyzing the information about unitary operations and wavefunctions. Fifth, there is a close relationship between quantum circuits and tensor networks, which may be advantageous for the development of QML in the framework of tensor networks [66]. Considering all this, it is desirable to investigate the performance of QML on a variety of practical applications using real-world data sets, such as cheminformatics, materials informatics, and other practical machine learning tasks.

4. Conclusions

In the present work, we have developed our QML models designed for predicting the toxicity of 221 phenols (QSAR/QSTR modeling), using the framework of the quantum–classical hybrid algorithm. To our knowledge, this is the first practical application of QML for a nonlinear regression task using a real-world data set.

We have numerically investigated how the different encodings, the variational circuit architectures, and the redundant encodings using multiple qubits affected the performance of our QML models. In our particular application, angle encoding was found to be useful in terms of flexibility in data representation owing to its simplicity and high nonlinearity. Furthermore, the results suggest that our entangler-enhanced encodings provided more expressive power in data representation than the previous ones, implying that quantum correlation could be useful for the feature map representation of classical data. The numerical results also indicate that, in our QML models, encoding each component of the input vector into multiple qubits appeared to be helpful for improving the performance in comparison with encoding each component into single qubit. Repeated circuit layers with CNOT blocks in the variational circuit provided a computational speedup compared with the original variational circuit based on the time evolution of Ising Hamiltonian.

Our QML models performed significantly better than the MLR models (51% and 12% increases for the training and validation sets), suggesting that the QML models succeeded in nonlinear regression tasks. Moreover, our simulations indicate that our best QML models were comparable to those obtained by RBF networks, while improving the generalization performance. We have also mentioned several perspectives on QML models, from a more general standpoint. Further improvements would be needed for the developments of the encoding method and of the evaluation of the cost functions (post-processing). Exploring noisy simulations and experiments on the real quantum hardware would be important in order to improve the QML models. The present study opens up the possibility that QML could be used for various nonlinear regression tasks such as cheminformatics and other machine learning applications.

*Note added.* After the release of our preprint version, two independent works on data encoding were released very recently: one is the work by Schuld, Sweke, and Meyer [67] and the other by Goto, Tran, and Nakajima [68]. They investigated *parallel* and *sequential* scenarios in repeated data encodings from a mathematical point of view; our work focuses on a practical application of QML on a real-world data set, using multiple data encoding with entanglement that is suited for our particular case. Our encoding approach may also be useful, since the number of qubits and the circuit depth are still limited in the real near-term quantum hardware.

## Appendix A. Numerical results using 15 qubits

**Table A1.** Coefficients of determination for the training and the validation sets ($R^2_{\text{train}}$ and $R^2_{\text{val}}$) using 13 different encoder circuits with 15 qubits (three copies of the quantum states), the optimized number of layers in the variational circuit $L$ ($3 \leq L \leq 12$) and the scaling factor $f_{\langle Z \rangle}$ for the expectation value $\langle Z \rangle$.

| Encoder Circuit ID | $R^2_{\text{train}}$ | $R^2_{\text{val}}$ | $L$ | $f_{\langle Z \rangle}$ |
|---|---|---|---|---|
| M | 0.766 | 0.793 | 7 | 6.0 |
| M–M–CZ | 0.793 | 0.808 | 6 | 10.0 |
| M–A1–CZ | 0.843 | 0.828 | 10 | 10.0 |
| M–A2–CZ | 0.861 | 0.820 | 12 | 10.0 |
| M–M–CNOT | 0.786 | 0.814 | 4 | 10.0 |
| M–A1–CNOT | 0.855 | **0.839** | 12 | 10.0 |
| M–A2–CNOT | 0.867 | 0.836 | 10 | 10.0 |
| A1 | 0.833 | **0.839** | 4 | 8.0 |
| A2 | 0.832 | 0.835 | 5 | 8.0 |
| A1–A1–CZ | 0.837 | 0.818 | 3 | 10.0 |
| A2–A2–CZ | 0.803 | 0.832 | 4 | 10.0 |
| A1–A1–CNOT | 0.863 | 0.834 | 9 | 10.0 |
| A2–A2–CNOT | **0.873** | 0.838 | 10 | 10.0 |

## Appendix B. Performance and computational costs for the variational circuit

**Table B1.** Performance comparison of the three variational circuit units: the coefficients of determination for regression tasks.

| Variational circuit unit | Coefficients of determination | | | |
|---|---|---|---|---|
| | $x^2$ | $e^x$ | $\sin x$ | $|x|$ |
| Time evolution of Ising Hamiltonian | 0.997 | 0.997 | 0.997 | 0.971 |
| Circuit unit containing a CNOT block | 0.998 | 0.999 | 0.999 | 0.971 |
| Circuit unit containing a CZ block | 0.729 | 0.970 | 0.997 | 0.793 |

**Table B2.** Matrix size, computational costs, and required memory for the calculation of the Trotter operator matrix.

| Number of qubits ($n$) | Dim. of the Trotter operator matrix [a] | Relative computational costs [b] | Memory for the Trotter operator matrix [MB] [c] |
|---|---|---|---|
| 5 | 32 | 1 | 0.05 |
| 10 | 1,024 | 32,768 | 48 |
| 15 | 32,768 | 1,073,741,824 | 49,152 |
| 20 | 1,048,576 | 35,184,372,088,832 | 50,331,648 |

[a] Calculated as $2^n$

[b] Estimated as $2^{3n-3}$ (the computational cost using 5 qubits was set to 1)

[c] Estimated as $16 \times 2^{2n} \times 3 / 1024^2$

# Appendix C. Comparison between pure and hybrid QML models

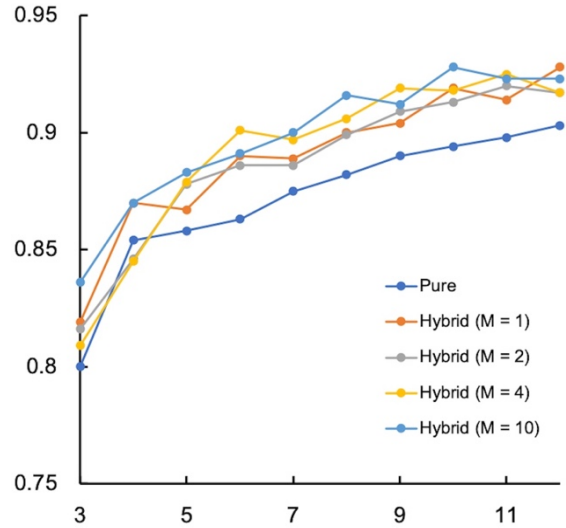

**Figure C1.** Comparison between pure and hybrid QML models. Coefficients of determination for the training set ($R^2_{\text{train}}$) as a function of the number of unit layers $L$ in the variational circuit $U(\boldsymbol{\theta})$. The results were obtained using the A2–A2–CNOT encoder with 10 qubits. In the hybrid QML models, the expectation values from $M$ qubits were fed into the evaluation of the loss function. The scaling factor $f_{\langle Z \rangle}$ was set to 4.0 (we found that, in the hybrid QML models, the scaling factor greater than 4.0 resulted in an unstable performance).

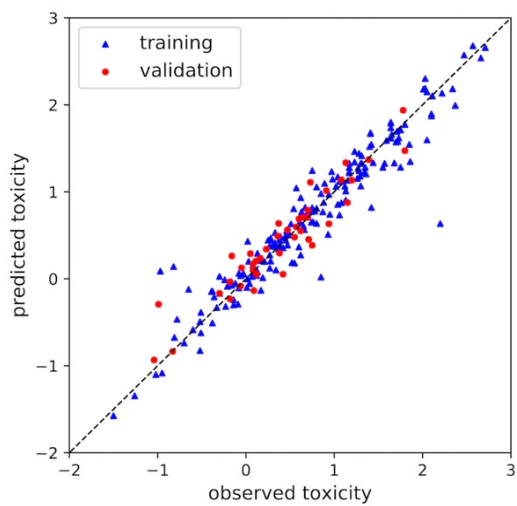

**Figure C2.** Plot for the observed vs. predicted toxicity obtained from the hybrid QML model ($M = 4$; $L = 10$) using the A2–A2–CNOT encoder with 10 qubits (blue triangle: training set; red circle: validation set) ($R^2_{\text{train}}$: 0.913; $R^2_{\text{val}}$: 0.886). Note that the plot is very similar to that obtained from pure QML model (using the single qubit measurement) in figure 6.

## Acknowledgment
We thank Seiya Sugo for useful discussions at the early stage of this work.

## Data availability
The data that support the findings of this study are available from the corresponding author upon reasonable request.

## ORCID iDs
Teppei Suzuki    https://orcid.org/0000-0001-7054-5493
Michio Katouda    https://orcid.org/0000-0001-7980-5386